# Field Testing of Residential Bidirectional Electric Vehicle Charger for Power System Applications

Shivam Saxena
*Dep. Electrical & Computer Eng.*
*University of New Brunswick*
Fredericton, Canada
shivam.saxena@unb.ca

Hany Farag & Khunsha Nasr
*Dep. Electrical Eng. & Comp. Science*
*York University*
Toronto, Canada
{hefarag,khunsha22}@yorku.ca

Leigh St. Hilaire
*Research & Development*
*Volta Research*
Toronto, Canada
leighsthilaire@voltaresearch.org

*Abstract*—Bidirectional electric vehicle (EV) charging is a technology that is gaining rapid popularity due to its ability to provide economic and environmental benefits to both EV owners and power system operators (PSOs). Using the EV as a flexible source of energy, an EV owner can provide power to homes/buildings, or even participate in grid services such as demand response and frequency regulation. However, there is a lack of real-world testing and validation for bidirectional charging technology, particularly in the residential segment. As such, this paper presents real-world field testing of a bidirectional EV charger deployed in a home. Control software is developed to dispatch the EV according to static setpoints, as well as automated load following, and its accuracy and responsiveness is reported on. The results of the testing with the charger and 2019 Nissan Leaf combination indicates a responsiveness of 6-8 seconds and accuracy of over 99%, which suggests feasible participation for applications such as load following, arbitrage, and demand response.

## I. Introduction

In an effort to reduce worldwide greenhouse gas emissions, there has been an acceleration of uptake with respect to electric vehicles (EVs), with over twenty countries mandating that all vehicle sales will be 100% EVs by 2030. [1]. This paradigm shift has motivated power system operators (PSOs), policy makers, and EV owners to propose new technologies to increase the usefulness of EVs. To address issues of power system congestion and associated degradation of power quality as a result of coincident EV charging, EV smart-charging programs have been successful, where PSOs can use EVs as a flexible load to reduce congestion at peak times [2]. However, studies show that even smart-charging programs may not be enough to handle the increase in peak demand caused by mass EV adoption, thus, necessitating expensive grid capacity upgrades by PSOs [3].

More recently, there has been great interest in bidirectional EV charging, where the onboard battery of the EV can be used as both a flexible load and a generation source. As such, the EV can be used to provide power to homes and buildings (V2H and V2B), and also to export power back to the grid (V2G) [4]. This level of functionality lends flexibility to both EV owners and PSOs, and results in a multitude of power system applications where bidirectional EV charging can be useful. For example, V2H can be used to provide power from the EV to specific circuits within a home, or to the entire home itself, especially in the case of power outages [5]. Similarly, V2B is effective in reducing peak demand charges of commercial buildings [6]. On the other hand, V2G provides the opportunity for EV owners to provide grid services to PSOs, whether participating in demand response or frequency regulation [7].

With the advent of power system decentralization and new markets opening to "behind-the-meter" participants, bidirectional charging is opening up new avenues in research [8]. In [9], the authors utilize EVs to perform arbitrage by charging their EVs at home during the night when electricity tariffs are lower, and discharging at work during the day when the tariffs are higher. The work in [10] develops a scheduling algorithm to minimize the time a home spends without power by using V2H. Meanwhile, the works in [11] and [12] develop optimization models to procure power from EVs for the services of demand response and frequency regulation considering the effects of battery degradation and revenue.

While the aforementioned works have contributed significantly to the state-of-the-art, it is worthwhile to note that these works have not used real-world data or experiences in their experimental results. As such, it is (a) difficult to determine the technical feasibility of bidirectional charging, especially pertaining to the speed and accuracy of power generation required by different power system applications; (b) unclear what methods are used to control the power output of the EV; and (c) challenging to ascertain what additional components are needed for bidirectional charging. These shortcomings prevent the mass uptake of bidirectional charging by the general public.

Motivated by these reasons, this paper presents a field evaluation of a residential, bidirectional enabled charger with respect to its speed of response, accuracy of power output, and subsequent suitability for power system applications, including demand response, frequency regulation, arbitrage, and load following. This paper contributes a novel attempt to disseminate real-world data and results from field tests, which to the best of the authors' knowledge, has not been shared before. Specifically, the charger has been deployed inside a real-world home and tested against real household appliances, from which the results can inform future algorithms, policies, and programs pertaining to bidirectional EV research.

The organization of the remainder of the paper is as follows. Section II provides a brief background of power system applications that bidirectional EVs can be utilized for, while Section III provides details of the interconnection and software-based control of the deployed charger. Section IV is devoted to discussing experimental results for the field testing of the charger, while Section V presents the conclusion of the paper.

## II. Background of Applications for Bidirectional EV Charging

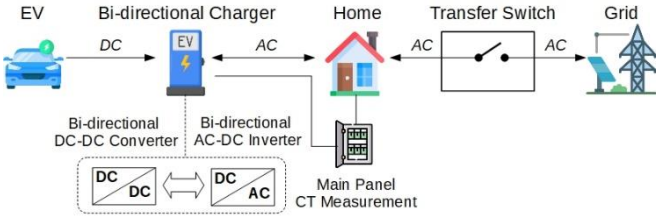

Fig. 1. Components of residential bidirectional EV charging.

## A. Components of Residential Bidirectional EV Charging

Fig. 1 illustrates the typical components used for residential Level 3 DC bidirectional EV charging, where the bidirectional charger houses a bidirectional inverter that couples with the EV battery to convert the DC energy of the battery to AC energy. It is worthwhile to mention that Level 2 AC bidirectional charging is possible if the EV itself is equipped with a bidirectional inverter [13], however, this paper focuses on Level 3 bidirectional charging. Nevertheless, the AC energy flows from the inverter and towards the home, where metering infrastructure is located at the point of common coupling between the home and the PSO-controlled grid. Using this measurement, the bidirectional charger (or external software) is able to export power from the EV equals to the load of the home to prevent back feeding the grid, if so desired. Furthermore, an automatic transfer switch is used to isolate the home from the grid during power outages, ensuring that the home is being powered only by the EV.

## B. Home Load Following

Used in the capacity of V2H, the EV is used to provide power to the loads of the home, both in grid-connected or off-grid scenarios. In a grid-connected scenario, the main motivation would be to utilize the EV to reduce electricity bills, particularly when running energy intensive loads, such as dryers, stovetops, and washing machines. In an off-grid scenario, or when there is a power outage, the EV would be used to keep power alive to critical loads, such as the heating source, internet, and fridge.

## C. Energy Arbitrage

Energy arbitrage opportunities can be realized when electricity tariffs fluctuate, typically as a function of time. For retail electricity rates offered to residential customers, a common tariff structure is known as time of use (TOU), which defines blocks of time where tariffs are at their highest (on-peak), and also at their lowest (off-peak). Thus, an EV owner may charge at off-peak hours, and then discharge at on-peak hours to capture daily revenue. Operating the EV in this manner generally leads to greenhouse gas emissions savings as well, since during on-peak times, PSOs activate "peaking power plants" to satisfy increased demand, leading to higher emissions [14].

## D. Demand Response

Demand response is typically used to reduce congestion within a power system during energy shortages. For the residential scenario, EVs can be used to generate power to reduce the load of their home, or, export any excess power to the power grid to further ease grid congestion. Demand response does not require very fast response from the EV, with timeframes in the minutes scale being acceptable [15].

## E. Frequency Regulation

Frequency regulation is typically used by PSOs to maintain balance between the supply and demand of power, which stabilizes the grid frequency. Power can be injected or withdrawn to maintain the balance, where PSOs send regulation signals, or setpoints, to participating entities to modulate their power output accordingly, in the timeframe of 2-6 seconds [16]. Bidirectional EVs could thus be a suitable candidate to provide frequency regulation services in aggregate if the responsiveness is indeed within 2-6 seconds.

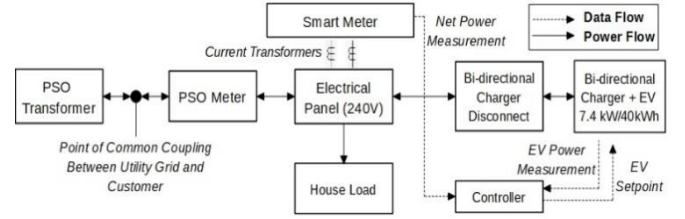

Fig. 2. Single line diagram of deployed charger within home.

## III. DEPLOYMENT OF RESIDENTIAL BIDIRECTIONAL CHARGER

### A. Physical Interconnection With PSO Grid

The single line diagram in Fig. 2 shows the physical interconnection and deployment of the bidirectional charger under test, within a home located in Ontario, Canada. The charger is rated for +/- 7.4 kW, and used with a 2019 Nissan Leaf SV that has 40 kWh battery capacity. In order to meet connection approval from the local PSO, a disconnect is deployed outside the home for the utility to turn off the charger during power outages. As seen in Fig. 2, a smart meter is deployed near the main electrical panel of the home, which uses current transformers within the panel to take net power measurements of the entire home. Also seen in Fig. 2 is an external software-based controller that is developed to take measurements from the smart meter and bidirectional charger in real-time, while also dispatching the charger with specific setpoints in units of kW. It is worth noting that the charger is not capable of islanding during a power outage, and as such, a transfer switch is not deployed within the home.

### B. Software Control of Bidirectional Charger

Both the smart meter, manufactured by Accuenergy (Acuvim-L series), and the charger allow external software to read/write values to these devices using the Modbus protocol. As such, both the meter and the charger expose specific registers, which the external software can use to toggle certain functionalities. For the smart meter, the registers of interest are the readings of the active power measurement and current flowing to/from the home. For the charger, the registers of interest are the ability to control the charger remotely (from external software), the ability to stop/start charging or discharging, the setting of a setpoint in kW, the reading of the power being delivered/received from the EV, and the state of charge (SOC) of the EV battery.

To control the power flow of the EV, particularly in relation to the loads of the home, a distributed energy resources management system (DERMS) was used, developed by Hero Energy and Engineering, entitled Harmonize [17]. The DERMS software provides control of energy resources for power flow optimization, however, it was expanded to develop libraries that facilitate the measurement and control of the bidirectional charger. Thus, the DERMS is responsible for reading relevant measurements from the smart meter and the charger at a configurable sampling rate, synchronizing the timing of the measurements, allowing both manual and automated control of the EV, and logging the data to disk.

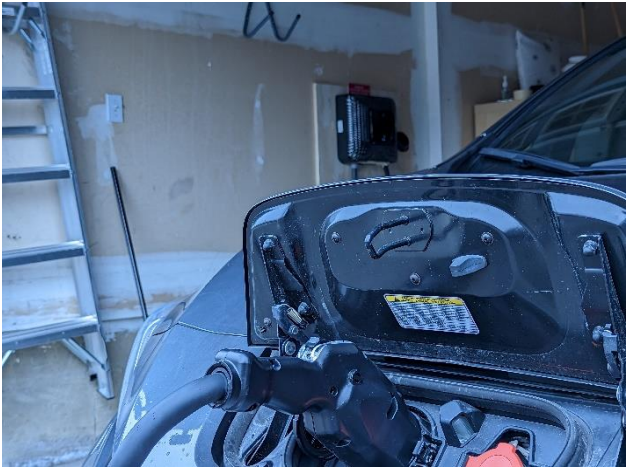

Fig. 3. Snapshot of EV and deployed bidirectional charger.

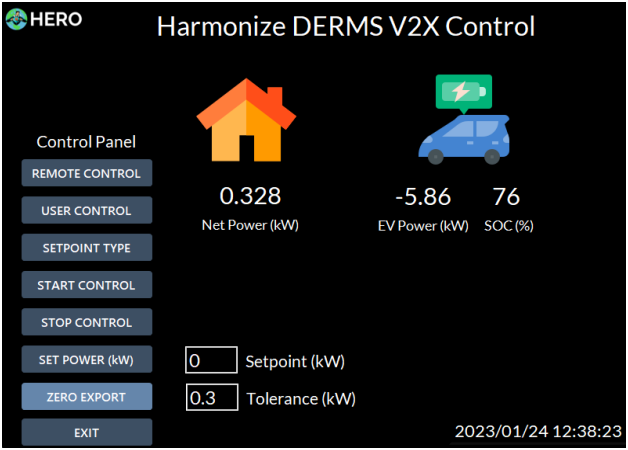

Fig. 4. Software controller dispatches EV to meet house loads in real-time.

A snapshot of charger and EV combination (Nissan LEAF 2019) is shown in Fig. 3, while a screenshot of the DERMS software is shown in Fig. 4, where functionalities to control the EV are provided in the Control Panel. For manual control, the user may select the "Set Power (kW)" button, and enter a value within the "Setpoint (kW)" textbox to send the setpoint to the charger. On the other hand, for automated control, the user may press the "Zero Export" button, which enables the EV to deliver just enough power to meet the loads of the home. As shown in Fig. 2, the DERMS receives measurements of the net power measurement of the home from the smart meter, as well as the EV power measurement from the charger, from which the setpoint is calculated and sent to the charger. The equation to realize the setpoint is as follows:

$$EV_{SET}(t) = P_{NET}(t-1) - P_{EV}(t-1) - \partial \quad (1)$$

where, $EV_{SET}$ is the EV setpoint, $P_{NET}$ is the net power of the home, $P_{EV}$ is the power met represents time, and α represents a tolerance factor, in kW, used as an offset to ensure the EV does not back feed the grid. The realization of (1) can be seen in Fig. 4, where the net power of the home is 0.328 kW, which is quite close to the tolerance set of 0.3 kW, while the EV is

The authors would like to thank the Natural Resources Canada Zero Emissions Vehicle Awareness Initiative (ZEVAI) for funding this work.

TABLE I. STEP TEST RESULTS

| Setpoint(kW) | EV Power (kW) | Time (s) | Error (%) |
|---|---|---|---|
| 7 (START) | 6.29 | 109.32 | 10.10% |
| 7 to 6 | 5.98 | 5.07 | 0.18% |
| 6 to 5 | 4.98 | 6.20 | 0.24% |
| 5 to 4 | 3.99 | 7.25 | 0.25% |
| 3 to 2 | 2.99 | 6.94 | 0.18% |
| 2 to 1 | 1.99 | 7.42 | 0.28% |
| 1 to 0 | 0.99 | 6.24 | 0.47% |
| 0 | 0.00 | 3.15 | 0.00% |
| 0 to -1 | -0.99 | 6.78 | 0.11% |
| -1 to -2 | -2.00 | 6.75 | 0.04% |
| -2 to -3 | -3.00 | 7.10 | 0.00% |
| -3 to -4 | -3.99 | 7.13 | 0.04% |
| -4 to -5 | -4.99 | 6.55 | 0.04% |
| -5 to -6 | -5.99 | 6.49 | 0.07% |
| -6 to -7 | -6.05 | 2.791 | 13.51% |

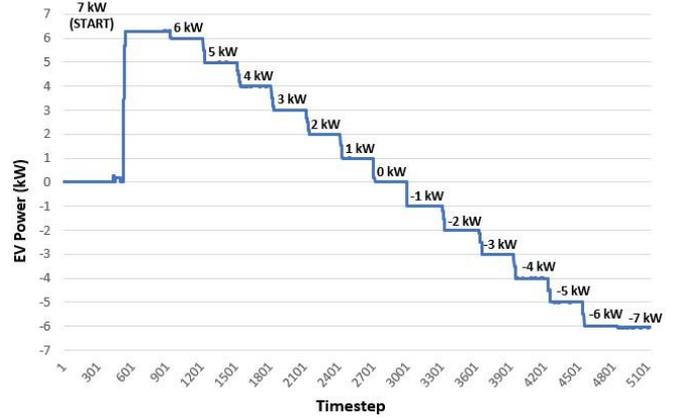

Fig. 5. Step test for bidirectional charging.

set to discharge 5.86 kW from its battery to meet the loading of the home.

## IV. EXPERIMENTAL RESULTS

This section presents experimental results to evaluate the speed, accuracy, and responsiveness of the power transfer of the deployed residential bidirectional charger, thereby enabling an evaluation with respect to its applicability to the power system applications discussed in Section II.

### A. Step Test

The step test involves setting setpoints on the charger in steps of 1 kW to evaluate its speed and accuracy. In this test, the sampling data acquisition rate from the meter and charger was set at 5 Hz, and setpoints were set in decrements of 1 kW every minute (or 300 timesteps). The response speed is calculated as the duration of time between the time the setpoint command was sent, and the first occurrence of the EV power measurement reaching the setpoint, while the accuracy is calculated as the percentage difference between the setpoint and the average EV power when the setpoint and arrives at at steady state. Three trials were repeated of this test, and the average set of results can be seen in Table I and Fig. 5.

TABLE II. SWEEP TEST RESULTS

| Setpoint(kW) | EV Power (kW) | Time (s) | Error (%) |
|---|---|---|---|
| -6 | -5.99 | 8.74 | 0.07% |
| 6 | 5.98 | 7.32 | 0.18% |
| -6 | -5.99 | 8.40 | 0.07% |
| 6 | 5.99 | 9.52 | 0.07% |
| -6 | -5.99 | 8.90 | 0.07% |
| 6 | 5.99 | 9.54 | 0.07% |

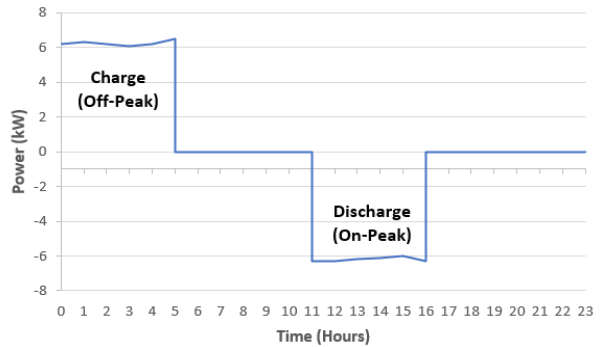

Fig. 7. Arbitrage test based on electricity tariff time of use.

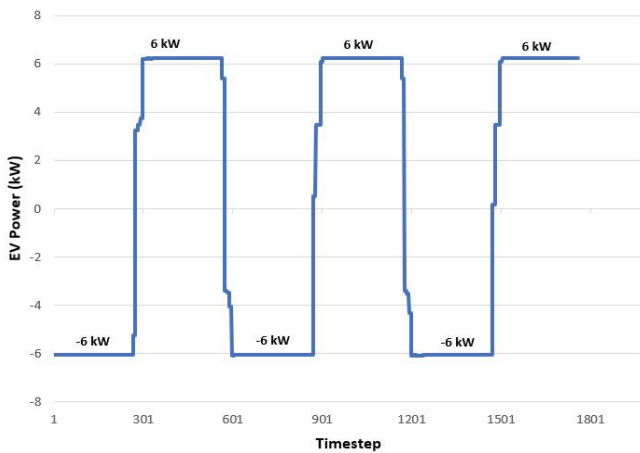

Fig. 6. Sweep test for bidirectional charging.

Three main observations can be made from Table I and Fig. 5. First, the starting setpoint takes a significant amount of time to reach when compared to other setpoints, with the starting setpoint being reached at 109 s versus an average of 6 s for the rest of the setpoints. This is presumably due to the initial negotiations between the charger and EV which occur on the first connection and subsequent charging requests [18]. Second, the charger has a maximum power transfer of +/- 6.3 kW when attempting to set setpoints at +/- 7 kW despite the nameplate claiming +/- 7.4 kW. This is because the charger is limited to a DC current input of 17 A, while the DC voltage of the EV battery measures at roughly 375V, resulting in a maximum power flow of 6.3 kW. Third, the accuracy of the charger is quite high when removing the samples of the +/- 7 kW setpoint, with an average error of less than 0.15 %. The step results indicate that the charger may work well for power system applications with less stringent timing requirements, such as demand response, load following, and arbitrage.

### B. Sweep Test

The sweep test is executed in a similar fashion to the step test, with setpoints being issues from +6 kW to – 6 kW. As seen in the results shown in Table II and Fig. 6, the setpoint accuracy provided by the charger is extremely high with an average error of less than 0.1%, however, it takes an average of 8.74 s to move through the entire allowable range. Recalling that the average frequency regulation signal is between 2-6 s [16], improvements in the response time are required before this charger can provide the frequency regulation service.

### C. Energy Arbitrage

The arbitrage test involves charging the EV at off-peak hours, and discharging the EV at on-peak hours. In Ontario, the 2022 retail electricity tariffs are $0.082/kWh for off-peak, and $0.17/kWh for on-peak, where the summer off-peak hours are between 19:00 and 7:00, while the on-peak hours are between 11:00 and 17:00. The results of the arbitrage test can be seen in Fig. 7, where the EV shows steady charging and discharging during off-peak and on-peak periods, respectively, potentially earning a total of $3.61 for the day. Note that Ontario regulations do not currently allow incentives for EVs to export energy to the grid. Nevertheless, the steadiness of the power discharge during the on-peak hours suggests that the charger would be able to participate in applications involving demand response, especially since the responsiveness required is more on the scale of minutes rather than seconds.

### D. Load Following with Zero Export

In this test, the EV is dispatched to follow the loads of the home. To this end, several common household appliances, such as the dishwasher, dryer, and washing machine are turned on during a test duration of approximately five hours to test the responsiveness of the EV and charger in load following. To minimize back feeding the grid, a tolerance of 0.1 kW is set as per (1), while all other loads within the homes are turned off, leaving a base load of approximately 0.13 kW throughout the test.

The results from the test can be seen in Table III, as well as Figs. 8 and 9, where the EV power mirrors the overall house load as per (1). In addition, as seen in Fig. 8, the net power of the home hovers close to the 0.1 kW mark throughout the majority of the test, except for periods for when the dryer is on. This is due to the fact that the dryer's power consumption changes very quickly, within 1-2 seconds, while the average response time of the EV is around 6 seconds (as found in the step test). Thus, Fig. 8 shows that the plot of the EV lags behind the plot of the net house power, thereby leading to periods of time where the EV cannot account for all house loads, resulting in the net house power being greater than 0.1 kW. Nevertheless, accounting for the tolerance of 0.1 kW, the EV is able to supply 4.37 kWh of shiftable house load of 6.41 kWh, or 68%. A plot of the SOC of the EV is also shown in Fig. 9, where the SOC drops from 85 to 66 during the test, with most of the decline occurring when the dryer is on.

TABLE III. LOAD FOLLOWING RESULTS

| Item | Duration (min) | Energy (kWh) |
|---|---|---|
| Kettle | 4 | 0.07 |
| Washing Machine | 57 | 0.25 |
| Dryer | 80 | 4.35 |
| Microwave | 2 | 0.02 |
| Dishwasher | 83 | 1.08 |
| Base Load | 303 | 0.65 |
| Tolerance | 303 | 0.505 |
|  | **TOTAL** | **6.91** |
| EV | 303 | -4.37 |
| Net House | 303 | 2.53 |
| House Load | 303 | **6.91** |

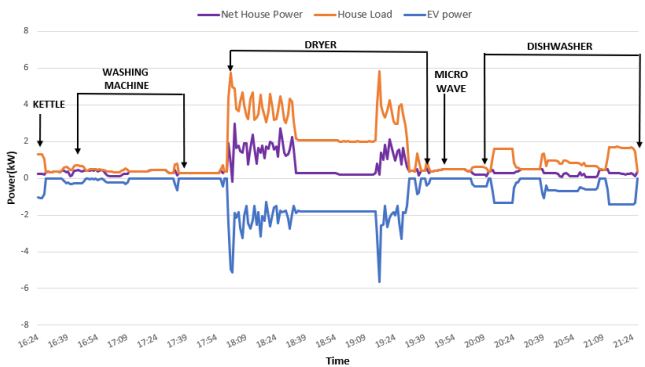

Fig. 8. Power transfer during load following test.

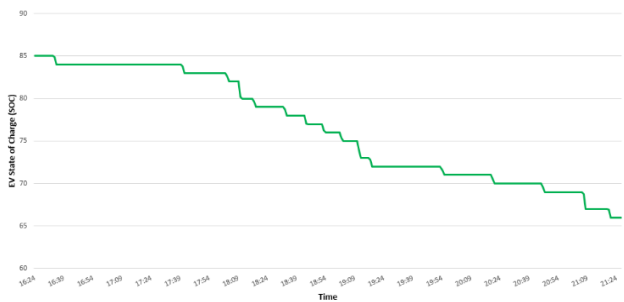

Fig. 9. EV SOC during home backup and zero-export test.

## V. CONCLUSION

This paper presents the real-world field testing of a residential bidirectional EV charger with a view towards evaluating its suitability for applications of load following, arbitrage, demand response, and frequency regulation. The charger was deployed to a home in Ontario, Canada, and outfitted with a smart meter to provide measurements of the net energy flow. DERMS software was enhanced to monitor and control an EV in real-time in response to house loads. Field tests with the charger and 2019 Nissan LEAF found that charger/EV combo responds to setpoints within 6-8 seconds, thereby being appropriate for applications of arbitrage, demand response, and home backup, while requiring improvement before being provisioned for frequency response.